\documentclass[aps,twocolumn,superscriptaddress]{revtex4-1}
\usepackage{times}
\usepackage{amsmath,amssymb,mathrsfs,physics}
\usepackage{graphicx}
\usepackage{stmaryrd}
\usepackage{amsmath,amsfonts}
\usepackage{color}

\begin{document}

\title{Null-eigenvalue localization of quantum walks on real-world complex networks}
\author{Ruben Bueno}
\email{ruben.bueno@espci.org}
\affiliation{\'{E}cole sup\'{e}rieure de physique et de chimie industrielles de la Ville de Paris, ESPCI Paris, PSL Research University\\
10, rue Vauquelin, 75231 Paris Cedex 05, France}
\affiliation{Institute of Industrial Science, The University of Tokyo, 4-6-1 Komaba, Meguro, Tokyo 153-8505, Japan}
\author{Naomichi Hatano}
\email{hatano@iis.u-tokyo.ac.jp}
\affiliation{Institute of Industrial Science, The University of Tokyo, 5-1-5 Kashiwanoha, Kashiwa, Chiba 277-8574, Japan}

\date{\today}

\begin{abstract}
First we report that the adjacency matrices of real-world complex networks systematically have null eigenspaces with much higher dimensions than that of random networks. 
These null eigenvalues are caused by duplication mechanisms leading to structures with local symmetries which should be more present in complex organizations. The associated eigenvectors of these states are strongly localized.
We then evaluate these microstructures in the context of quantum mechanics, demonstrating the previously mentioned localization by studying the spread of continuous-time quantum walks.
This null-eigenvalue localization is essentially different from the Anderson localization in the following points: first,
the eigenvalues do not lie on the edges of the density of states but at its center; second,
the eigenstates do not decay exponentially and do not leak out of the symmetric structures.
In this sense, it is closer to the bound state in continuum.
\end{abstract}


\maketitle

\section{Introduction}

Complex networks define relations between entities like atoms, proteins, or even humans~\cite{intro1, intro2, intro3}, as in acquaintance networks~\cite{zk} and the World Wide Web~\cite{www}.
They are at the crossroad of several disciplines including physics, mathematics, biology, and even social sciences. 
The development of computer science has allowed researchers to analyze and experiment in details the structure of these systems. 
Mathematical models either of regular lattices or of random graphs appeared to be insufficient~\cite{intro4} to explain complex characteristics that real-world networks exhibit, including the small-world phenomenon~\cite{milgram} and scale-free property~\cite{strogatz}.
The former phenomenon refers to the finding that the shortest path between arbitrary two nodes is much shorter in complex networks than in random graphs.
The latter property refers to the discovery that the histogram with respect the number of links attached to each node has a power-law behavior for many real-world networks.
In fact, the term ``complex networks" is often used only to mark the difference of real-world networks of wide variety from random graphs considered in mathematical graph theory.

Let us note that most of the approaches to complex networks focus on global features of networks as reviewed above.
In contrast, we here focus on local symmetries that typically appear in real-world complex networks.
More specifically, we analyze networks in terms of the eigenstates of their associated matrix, namely the adjacency matrix~\cite{intro5}, and find that many of the eigenstates belong to the null eigenspace and are localized in locally symmetric structures.

This localization has distinct features from the celebrated Anderson localization~\cite{localander}.
The latter takes place typically near the edges of the spectrum, while the present one, which we will refer to as \textit{the null-eigenvalue localization}, happens exactly at the center of the spectrum.
This is in striking contrast to random tight-binding models with the chiral symmetry, whose state at the center of the spectrum, namely the zero-energy mode, has a divergent localization length~\cite{zerostate1,zerostate2,zerostate3,zerostate4,zerostate5} with all other states being localized because of the Anderson localization.

The eigenstates under the null-eigenvalue localization also looks quite different from the typical eigenstates under the Anderson localization. 
The former are strictly caged on a set of nodes, that is, the amplitude of the former is finite only on a restricted number of nodes, while the latter typically decay exponentially;
admittedly, there are cases of caged eigenstates of the latter, but they are quite exceptional.
We claim that the caged eigenstate at the center of the spectrum has more similarity to the phenomenon of the bound state in continuum~\cite{BIC1,BIC2,Ordonez06,Tanaka07}.

We finally analyze the effect of the null-eigenvalue localization in the context of quantum mechanics by studying the time evolution of a continuous-time quantum walk (CTQW)~\cite{QW1, CTQW1, CTQW2, CTQW3, review1, review2, review3}.
We will demonstrate that the infinite-time average of the transition probability is localized inside the cage of the null-eigenvalue state.
We note that similar analyses have been previously made on artificial hierarchical networks~\cite{CTQW2,CTQW3}, but the results were not connected to the null-eigenvalue localization that typically happens in real-world networks.

Apart from localization of CTQW on complex networks, we note that localizations were documented in the case of discrete-time quantum walks (DTQW) on regular lattices~\cite{DiscreteLoca1, DiscreteLoca2}. 
The degeneracy of eigenvalues was necessary for the localization in this specific context, because superposition of eigenvectors with different wave numbers produce a localized state.
A recent study~\cite{Mukai20} explored DTQWs on complex networks, and found that a quantum walker tends to be localized on a few nodes if the time-evolution unitary matrix has strong degeneracies, while it tends to be localized in a community if the operator does not.
In contrast, we will see below that the localization of the CTQW on \textit{irregular} networks would occur without degeneracy.
In our context, the degeneracy is not a cause of localization but a consequence of the fact that there are a large number of localized states of null eigenvalues all over the real-world network.

This paper is organized as follows. First in Sec.~\ref{sec2}, we explore the concept of nullity in graph theory and complex networks.
We stress that the adjacency matrix of real-world complex networks typically has an eigenvalue spectrum that is very distinct from the one of the random matrix theory, in that the former have high degeneracy of null eigenvalues.
In Sec.~\ref{sec2.5} we understand the null-eigenvalue degeneracy in the light of quantum mechanics. We claim  that locally symmetric structures of typical complex networks produce geometrical constriction of wave functions, caging the null eigenvectors. 
Section~\ref{sec2.6} confirms the null-eigenvalue localization in terms of the inverse participation ratio.
We finally reveal in Sec.~\ref{sec3} the impact of these null eigenstates on the CTQW numerically in real-world examples. Section~\ref{sec4} concludes the paper.

\section{High null-eigenvalue degeneracy}
\label{sec2}
A graph (network) $\mathcal{G}=(V,E)$ is a combination of vertices (nodes) $V\in\llbracket 1;N \rrbracket$ and edges (links) $E\in\llbracket 1;N \rrbracket^{2}$. 
Its adjacency matrix $A$ is given by
 \begin{align}\label{eq10}
A_{ij}=
\begin{cases}
1 & \mbox{for } (i,j)\in E,  \\ 
0 & \mbox{otherwise.}
\end{cases}
\end{align}
We focus on undirected networks so that the adjacency matrix may be kept real, symmetric and diagonalizable with eigenvalues $\{\lambda_\mu|\mu=1,2,\cdots,N\}\in\mathbb{R}^{N}$.
The corresponding eigenvectors $\{|\phi_\mu\rangle\}$ satisfy the orthonormality
$\langle\phi_\mu|\phi_\nu\rangle=\delta_{\mu\nu}$.
This allows us to establish a bijection between a network and the spectral properties of its corresponding matrix.
Therefore, any topological information such as communities and hierarchies~\cite{Mukai20,fortunato, EHB, hatano, estradabook, statis1, statis2} should be contained in the spectrum.

\begin{figure*}
\centering
\includegraphics[width=0.3\textwidth]{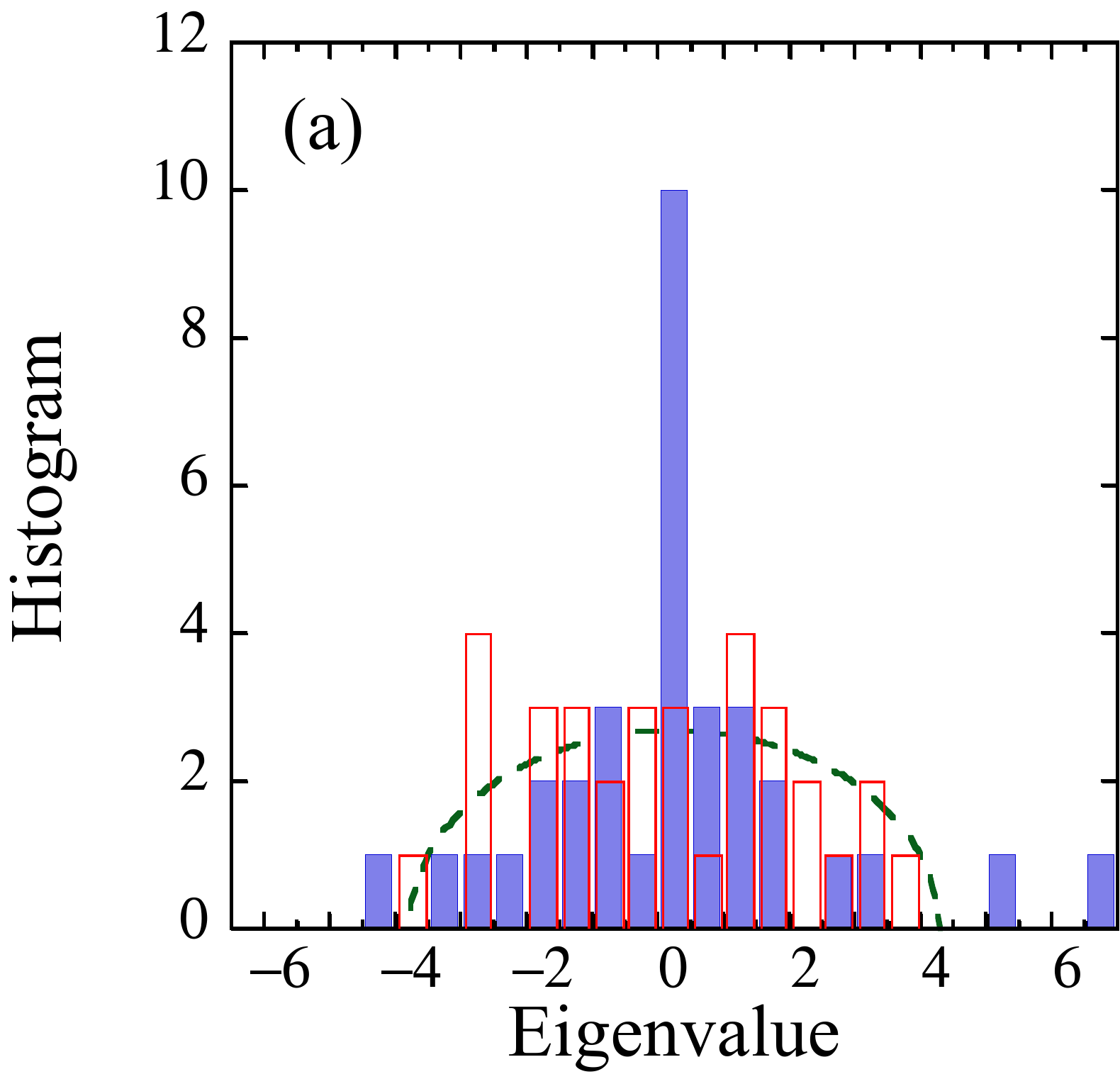}
\hspace{0.15\textwidth}
\includegraphics[width=0.3\textwidth]{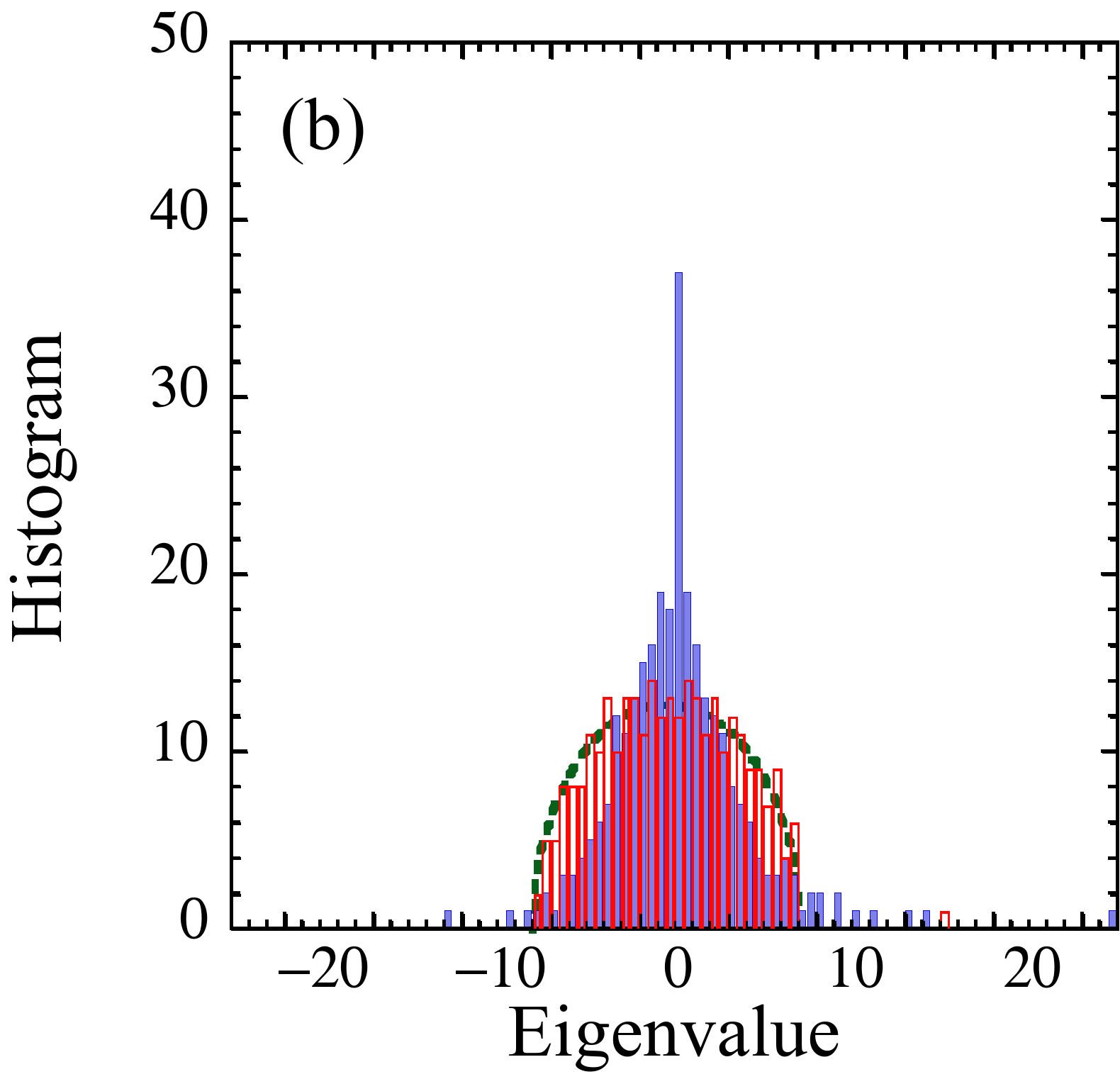}
\\[\baselineskip]
\includegraphics[width=0.3\textwidth]{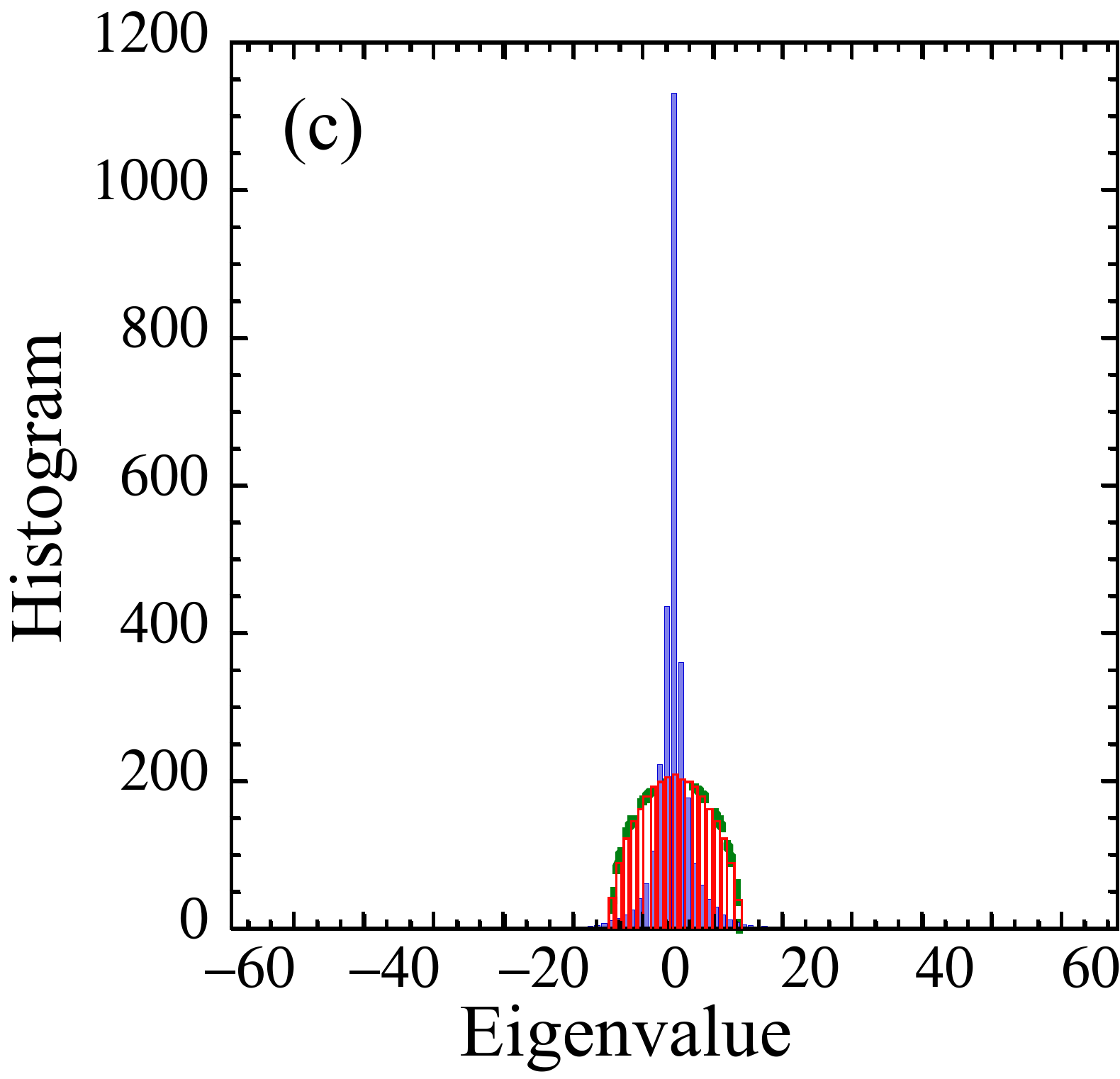}
\hspace{0.15\textwidth}
\includegraphics[width=0.3\textwidth]{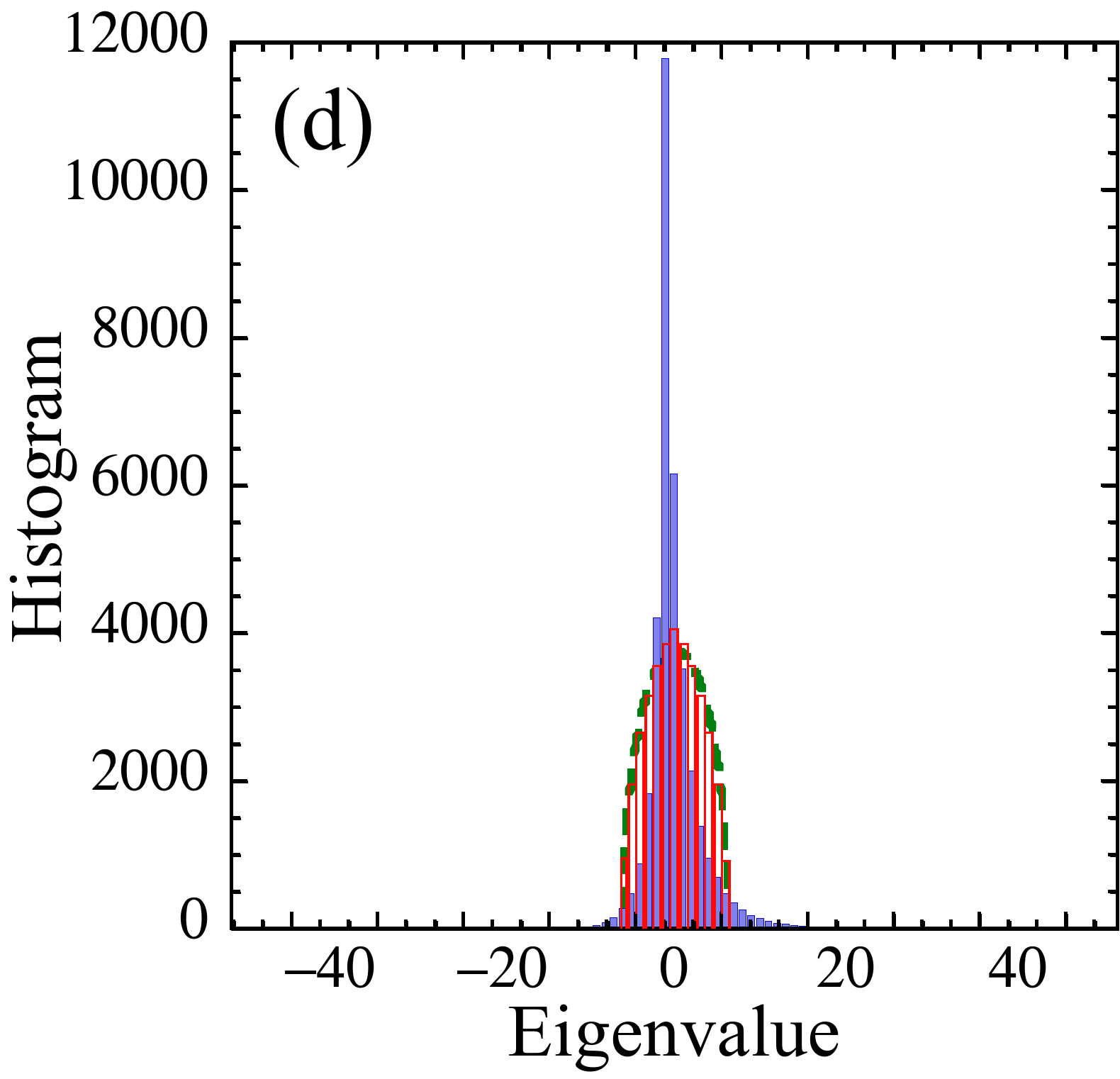}
\caption{Eigenvalue distributions of the adjacency matrices computed from two networks and their random-network counterparts: 
(a) Zachary's karate-club network~\cite{zk}; 
(b) neural network of \textit{c.\ elegans}~\cite{neuralnetwork}.
(c) airline connections between airports in the U.S.~\cite{airline}; 
(d) collaboration network in preprints on the condensed-matter archive at www.arxiv.org~\cite{condmat}.
In each panel, the shaded columns indicate the eigenvalue histograms (\textit{i.e.}\ the number of eigenvalues in each bin) of the complex network, while the open columns the ones of their random-network counterparts with the same numbers of nodes and links. The counterpart did not have any null eigenvalues in both cases. We build the random networks according to the following 5 steps: (i) generation of a random number in the range of $1$ to $N$ and choose a node; (ii) generation of another random number in the range of $1$ to $N$ and choose a node; (iii) If the node chosen in (ii) is the same as the node chosen in (i), repetition of (ii) until they are different; (iv) connection of the two nodes with a link; (v) repetition of the process from (i) to (iv), rejecting multiple links, until having links of a predetermined number.
The broken curve indicates the semi-circle law according to Ref.~\cite{bauer}, which reproduces the histogram for the random network well.}
\label{data_networks_raw}
\end{figure*}

We exemplify in Fig.~\ref{data_networks_raw} the eigenvalue distributions of four complex networks~\cite{zk,neuralnetwork,airline,condmat}.
We clearly see a prominent peak at $\lambda_{\mu}=0$.
On the other hand, their random-network counterparts (a random network with the same number of nodes and links, $N$ and $K$) tend to yield eigenvalue distributions that follow Wigner's semi-circle law of the random-matrix theory~\cite{wigner,bauer}, which is a semi circle of the range $|\lambda_\mu|\leq 2\sqrt{p(1-p)N}$ except for a large eigenvalue, where $p=2K/[N(N-1)]$ is the linking rate.
We also notice other peaks around integer eigenvalues, but we focus on the null eigenspace hereafter.

\begin{table}
\caption{Details of the null eigenspaces of four networks. 
The linking rate $p$ is the number of links $K$ divided by $N(N-1)/2$ with $N$ the number of nodes in the largest cluster.
We took only the largest connected subgraph; in other words, we excluded any smaller subgraphs including isolated nodes.
We also replaced all directed links with undirected ones. We then estimated the number of null eigenvalues by computing the rank of thus-created adjacency matrix.
We additionally generated a random-network counterpart for each complex network by preparing a connected graph with the same numbers of nodes and links. 
The counterpart did not have any null eigenvalues in every case.
(\# refers to ``the number of.")
}
\label{comparison_table}
\begin{tabular}{lrrrr}
\hline
& Karate    & Neural               & Airline                   & Cond-mat\\
& \cite{zk} & \cite{neuralnetwork} & \cite{airline} & \cite{condmat}\\
\hline\hline
total \# nodes & 34 & 297 & 2939 & 40421 \\
\hline
\# nodes $N$ & 34 & 297 & 2905 & 36458 \\
in the largest cluster &&&&\\
\hline
\# links $K$ & 78 & 2162 & 30442 & 171735\\
in the largest cluster &&&&\\
\hline
\# links per node $K/N$ & 2.3 & 7.3 & 10.5 & 4.3 \\
\hline
linking rate $p$ & 13.9\% & 4.92\% & 0.72\% & 0.026\% \\
\hline
\# null eigenvalues & 10 & 15 & 808 & 1700 \\
fraction in $N$ & 29.4\% & 5.1\% & 27.8\% & 4.66\% \\
\hline
\end{tabular}
\end{table}

We counted numerically the null eigenvalues of four different networks in Table~\ref{comparison_table}.
(For these statistics and all analyses hereafter, we focused on the largest connected subgraph, removing any smaller subgraphs, which tends to have a higher rate of null eigenvalues, especially any isolated nodes, which yield trivial null eigenvalues.)
We then notice that the fraction of null eigenvalues in complex networks is significantly high, while their random-network counterparts did not have any null eigenvalues at all.

The peak at $\lambda_{\mu}=0$ has been reported in different fields such as condensed matter physics~\cite{alloy}, random-matrix theory~\cite{deltapeaks,kernel,sparse1,sparse2,sparse3,sparse4,sparse5,sparse6}, and various spectral analyses of random networks~\cite{spectra, deformed, dynamic,nullity} (although some of the previous analyses did not mention the removal of smaller clusters and isolated nodes) or spectral analysis of very specific networks ~\cite{vukadinovic, cantor}.
In random-matrix theory, this null-eigenvalue degeneracy was studied numerically~\cite{sparse1} and analytically, reaching the conclusion that it comes from the decrease in the rank of the matrix when increasing the sparsity~\cite{rankdecrease}. 
In Ref.~\cite{spectra}, they studied local tree-like networks and claimed that the degeneracy is due to dead-end vertices (We will show below that there are indeed many other structures that produce null eigenvalues).
A recent study looked into the details of the origin of the null-eigenvalue from a graph-theory perspective, and used these insights to understand how real-world networks grow~\cite{growth,growth2,growth3}. 
In the present study, we look into the null-eigenvalue degeneracy in terms of network structures, its localized properties, and its consequences in quantum mechanics.

\section{Quantum-mechaniccal interpretation of the null eigenvalues}
\label{sec2.5}

Null eigenvalues appear in the adjacency matrix under the following conditions~\cite{growth}:
\begin{description}
\setlength{\parsep}{0em}
\setlength{\parskip}{0em}
\setlength{\itemsep}{0em}
\setlength{\topsep}{0em}
\setlength{\partopsep}{0em}
\item[Complete duplication] if two rows (or columns) have exactly the same entries; 
\item[Partial duplication] if two or more rows (or columns) added together have exactly the same entries as some of the other rows (or columns); 
\item[Isolated nodes] if the network contains isolated nodes.
\end{description}
The last case leads to a row of null entries which trivially results in a null eigenvalue. We ignore this specific situation as we stated above.

\begin{figure}\centering
\includegraphics[width=0.48\textwidth]{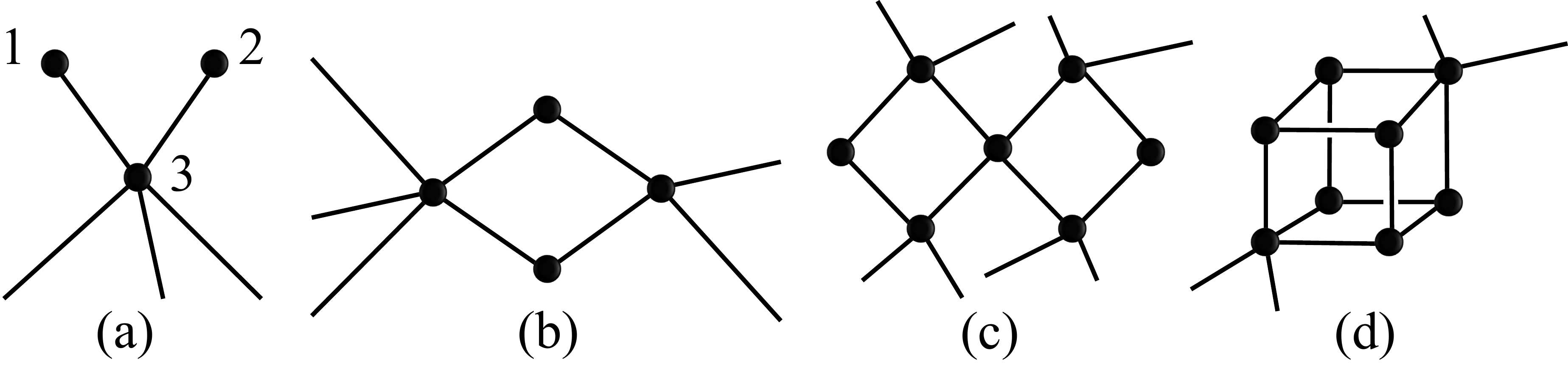}
\caption{(a) Two dangling nodes. (b--d) Other variations of complete duplications.
The one in (d) was taken from Ref.~\cite{alloy}, which is a study on alloy.
}
\label{trois}  
\end{figure}

To understand complete duplication intuitively, see Fig.~\ref{trois}(a), which exemplifies the simplest case between nodes 1 and 2. 
The corresponding adjacency matrix should take the form
\begin{align}\label{eq20}
A=
\bordermatrix{
& 1 & 2 & 3 & 4 & \cdots \cr
1 & 0 & 0 & 1 & 0 & \cdots  \cr
2 & 0 & 0 & 1 & 0 & \cdots   \cr
3 & 1 & 1 & 0 & * & \cdots \cr
4 & 0 & 0 & * & * & \cdots \cr
\vdots & \vdots & \vdots & \vdots & \vdots & \ddots \cr
},
\end{align}
where the integers outside the parentheses denote the node indices in Fig.~\ref{trois}(a).
The symbol~$\ast$ in Eq.~\eqref{eq20} can be either $0$ or $1$ depending on structures outside the dangling nodes. 
We can easily confirm that 
\begin{align}\label{eq30}
\frac{1}{\sqrt{2}}
(1, -1, 0, 0, \cdots)^T
\end{align}
is an eigenvector for the null eigenvalue.
The nullity comes from the fact that the first and second columns are equal to each other, which we mean by the complete duplication.

An intuitive understanding of this characteristic can be drawn in the context of quantum mechanics. The wave amplitude $+1/\sqrt{2}$ on the node $1$ and the wave amplitude $-1/\sqrt{2}$ on the node $2$ interfere with each other on the node $3$ and vanish, not leaking outside. The same argument applies to the other structures exemplified in Fig.~\ref{trois}(b--d) but with cancelation of different amplitudes.

This feature of localized wave function reminds us of the bound state in continuum~\cite{BIC1,BIC2,Ordonez06,Tanaka07}, which is a high-energy bound state formed not by a potential well but because of a geometrical constriction;
a wave function interferes with itself constructively inside a restricted area while destructively outside it; see Fig.~4 of Ref.~\cite{Tanaka07} for illustration. From there, we can predict that such a structure resulting from a complete duplication would lead to the localization of the wave function. 

We stress here again that the huge degeneracy of the null eigenvalue of localized states are quantitatively different from the one found in the studies of the discrete-time quantum walks (DTQW)~\cite{DiscreteLoca1, DiscreteLoca2,Mukai20}, as we commented in Introduction. 
In DTQW, the localization would not take place without the degeneracy;
only proper superposition of eigenvectors under the degenerate eigenvalue can produce localized states in DTQW.
In contrast, the present null-eigenvalue degeneracy emerges as a consequence of the fact that we have the situation of duplication (e.g. as in Fig.~\ref{trois}) all over the network.
Indeed, we could easily come up with a situation in which there is only one localized eigenstate with null eigenvalue.

In the case of partial duplication, a linear combination of a set of rows would have exactly the same entries as a linear combination of another set of rows. An eigenvector for this null eigenvalue would be a combination of wave amplitudes cancelling each others. 
The wave function will also vanish outside of the initial state but will be spread on more than two nodes. 
We will show an instance of the partial duplication later in Fig.~\ref{networksamples}(a).

\section{Inverse participation ratio (IPR) of the localized eigenstates}
\label{sec2.6}

\begin{figure}
\centering
\includegraphics[width=0.38\textwidth]{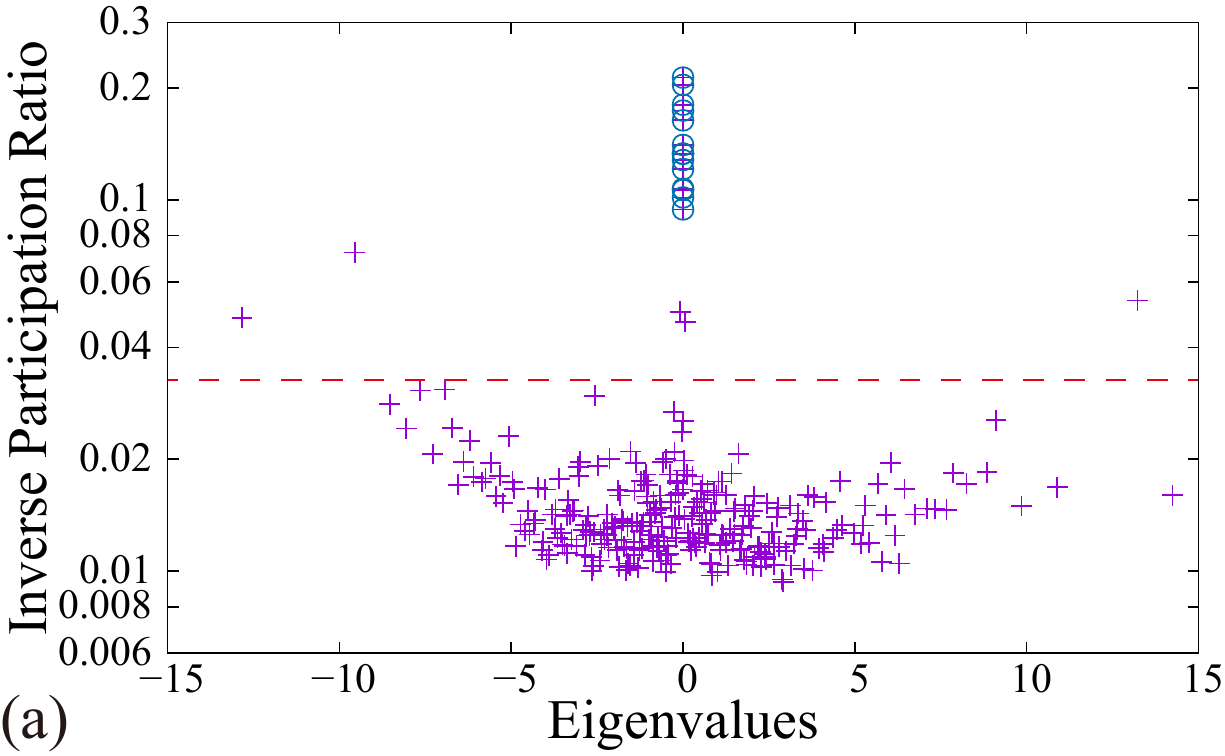}
\\[\baselineskip]
\includegraphics[width=0.38\textwidth]{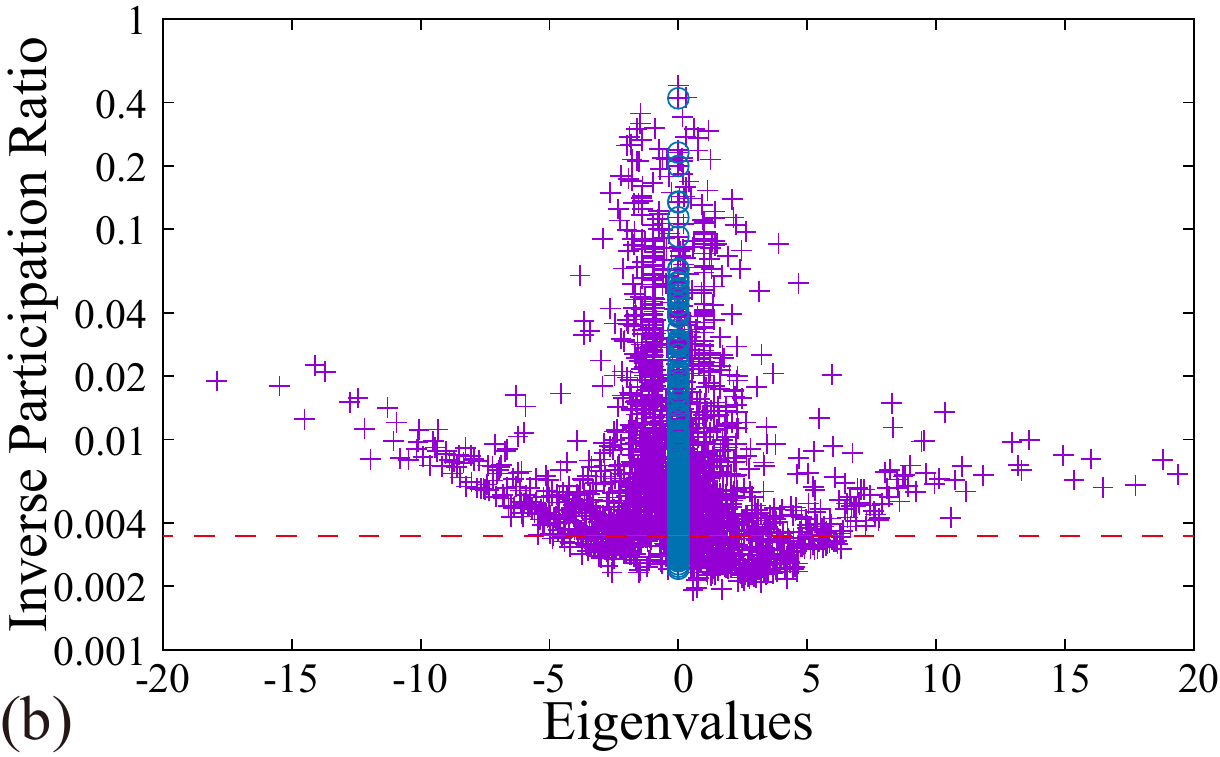}
\caption{The IPR of each eigenvalue of (a) the neural network of \textit{c.\ elegans}~\cite{neuralnetwork} (see Fig.~\ref{data_networks_raw}(b)) and (b) the U.S.\ airline network~\cite{airline}. (Details of the two networks are given in Table~\ref{comparison_table}).
In each panel, the crosses encircled indicate the null eigenvalues and the horizontal broken line indicates our criterion of localization, IPR$_\mu=10/N$.}
\label{IPR}
\end{figure}

The null-eigenvalue localization can be further demonstrated by the calculation of the inverse participation ratio (IPR) for a $\mu$th eigenvector $\ket{\phi_\mu}$,
\begin{align}\label{IPRdef}
\textrm{IPR}_\mu=\frac{\sum_{j=1}^N\langle j|\phi_\mu\rangle^4}%
{\left(\sum_{j=1}^N\langle j|\phi_\mu\rangle^2\right)^2},
\end{align}
which is greater for a more localized eigenstate.
Figure~\ref{IPR} shows IPR for each eigenstate of two networks. 
This markedly illustrates that the null eigenvectors constitute a substantial part of the spectrum, and have relatively large values of IPR in the center of the eigenvalue distributions. We will call this effect the ``null-eigenvalue localization" because it is fundamentally different from the Anderson localization, which tends to happen for eigenstates near the edges of the spectrum. The IPR indeed increases near the edges too in Fig.~\ref{IPR}, which can be understood in the context of the Anderson localization.

More quantitatively, the IPR in Eq.~\eqref{IPRdef} would take the value of $1/N_1$ if the eigenvector has amplitude equally over $N_1$ pieces of nodes and vanishes at the other nodes.
Reversing the logic, we can define an effective number of nodes over which the eigenvector resides, by inverting the IPR.
For the sake of the argument, let us tentatively define localization as the case in which the effective number of nodes is less than one tenth of the total number of nodes, that is when $N_1<N/10$, or IPR$_\mu >10/N$.

For the neural network with $N=297$, whose IPR is depicted in Fig.~\ref{IPR}(a), the tentative localization criterion is IPR$_\mu>10/297\simeq 0.033$, which is indicated by a horizontal broken line in the figure.
All 15 eigenstates degenerate in null eigenvalue clear this criterion by far with their IPRs ranging from $0.101$ up to $0.213$.
The effective number is about $2.9$ for the most localized eigenstate, which has a null eigenvalue.
On the other hand, the eigenvector with the eigenvalue $2.9167\cdots$ has the lowest IPR, which is about $0.0093$ and falls below the localization criterion; 
the effective number of nodes is about $107$. 
Meanwhile, as an example of states near the edge of the spectrum, for the state with the eigenvalue $-9.548\cdots$, the IPR is about $0.0483$, which clears the localization criterion, and the effective number of nodes is about $13.9$. 
This eigenstate is less localized than the states of the null-eigenvalue localization but still satisfies our tentative criterion of localization.

For the U.S.\ airline network, whose IPR is depicted in Fig.~\ref{IPR}(b), the tentative localization criterion is IPR$_\mu>10/2905\simeq 0.00344$, as is indicated by a horizontal broken line in the figure. 
The eigenstate with the highest IPR is one with null eigenvalue, and its IPR is about $0.4828$, whereas the eigenstate with the lowest IPR is one with the eigenvalue about $0.568$, and its IPR is about $0.00019$.
Meanwhile, the eigenstate of the lowest eigenvalue $-21.8\cdots$ has the IPR of about $0.0143$.
These data reveal that the null-eigenvalue localization and the Anderson localization happen in distinct ranges of eigenvalues.

\section{Consequences of the continuous-time quantum walk (CTQW)}
\label{sec3}
We now inspect a quantum-mechanical effect of the null-eigenvalue degeneracy.
We consider a quantum walk based on the adjacency matrix, that is, the time evolution of a quantum particle according to the Schr\"{o}dinger equation with the adjacency matrix as the Hamiltonian. We focus here on the continuous-time quantum walk (CTQW)~\cite{review1,review2,review3}, which is considered a promising model to describe coherent transport on complex networks. 
It is an equivalent of the tight-biding approximation in solid-state physics or the H\"uckel method in molecular orbital calculations. 

One needs to be aware that we will use the adjacency matrix to define our QW. However, in some other studies, researchers used the Laplacian matrix. We have confirmed that the characteristics we are analyzing are the same for both matrices. 

We span all the accessible Hilbert space by the states $|j\rangle=(0,\cdots,0,\stackrel{j}{1},0,\cdots,0)^T$ endowed with the node $j$ of the network. 
The probability that the particle starting from a node $k$ ends up at a node $j$ after time $t$ is given by the propagator 
\begin{align}\label{eq40}
\displaystyle
\pi_{jk}(t)
& = |\langle j|e^{-iAt}|k\rangle|^2.
\end{align}
Expanding Eq.~\eqref{eq40} with respect to the eigenstates $\{\phi_\mu\}$, we obtain diagonal and cross terms:
\begin{align}
&\pi_{jk}(t) = \sum_{\mu=1}^N\left|\langle j | \phi_\mu\rangle \langle \phi_\mu | k \rangle \right|^2
\nonumber\\ 
&+2 \sum_{\mu<\nu} \langle j | \phi_\mu \rangle \langle \phi_\mu | k \rangle \langle k | \phi_\nu \rangle \langle \phi_\nu | j \rangle \cos[(\lambda_{\mu} - \lambda_{\nu})t].
\label{propagatornode}
\end{align}  
Since we are interested in the localization of the particle, we consider the long-time average given by
\begin{align}\label{eq50}
\chi_{jk}=\lim_{T\to\infty}\frac{1}{T}\int_0^T\pi_{jk}(t) \mathrm{d}t.
\end{align}
Every term in the first summation in Eq.~\eqref{propagatornode} survives, while terms in the second one generally vanish except when the two eigenvalues are degenerate $\lambda_\mu=\lambda_\nu$.
We thereby obtain
\begin{align}\label{eq70}
\chi_{jk} &=  \sum_{\mu=1}^N \left| \langle j | \phi_\mu\rangle \langle \phi_\mu | k \rangle \right|^2
\nonumber\\
&+2 \sum_{\shortstack{$\scriptstyle \mu<\nu$ \\ $\scriptstyle \lambda_\mu=\lambda_\nu$}} \langle j | \phi_\mu \rangle \langle \phi_\mu | k \rangle \langle k | \phi_\nu \rangle \langle \phi_\nu | j \rangle.
\end{align} 

We can roughly estimate the relative magnitudes of the two summations in Eq.~\eqref{eq70}.
There are $N$ terms in the first one and  $N_{\lambda_\mu}(N_{\lambda_\mu}-1)$ terms in the second (including the factor 2), where $N_{\lambda_\mu}$ denotes the dimensionality of the eigenspace of $\lambda_\mu$. 
Therefore the second summation can dominate if $N_{\lambda_\mu}$ is a considerable part of $N$, which is indeed the case for $\lambda_\mu=0$ in Table~\ref{comparison_table}. 

\begin{figure}
\centering
\includegraphics[width=0.13\textwidth]{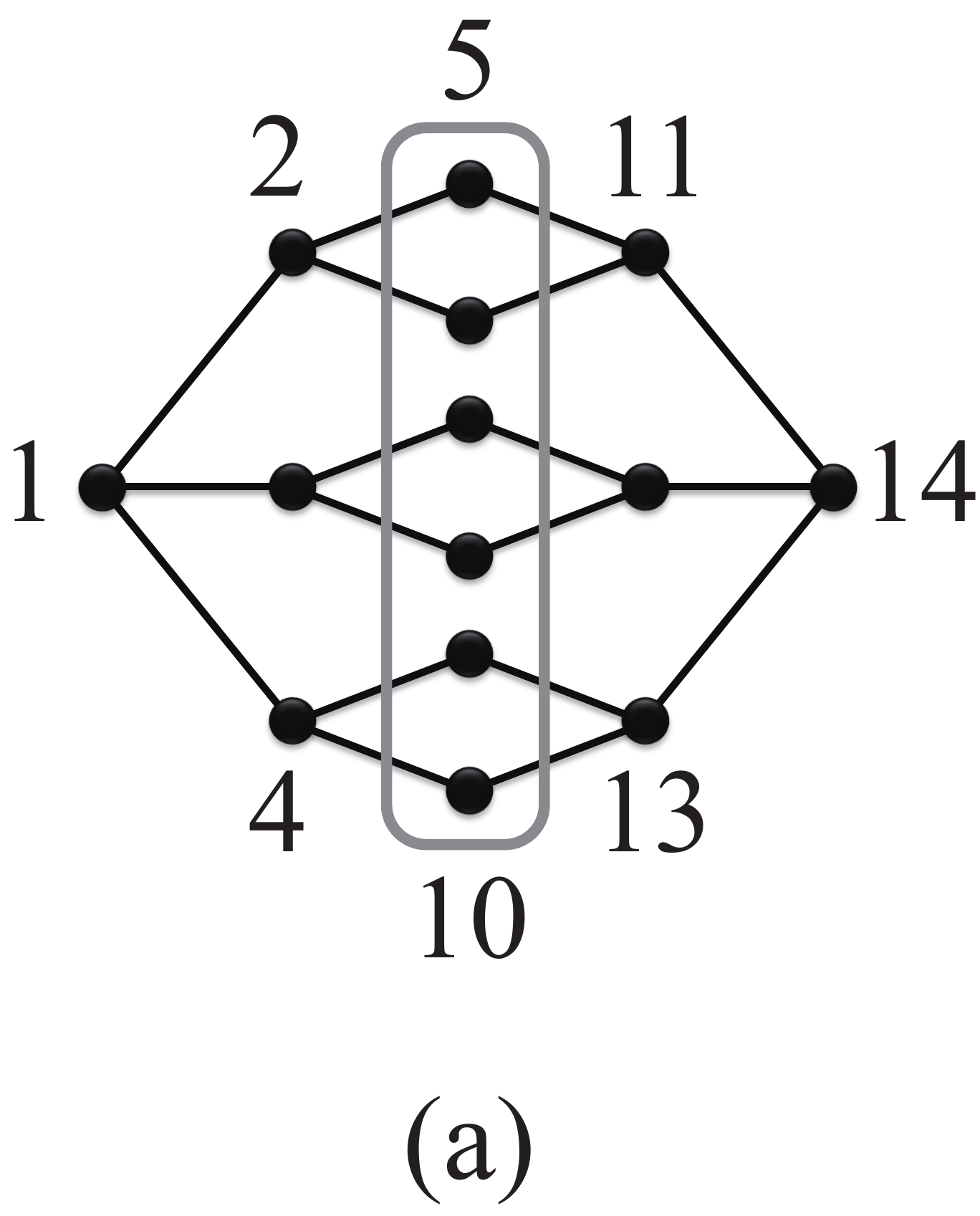}
\includegraphics[width=0.34\textwidth]{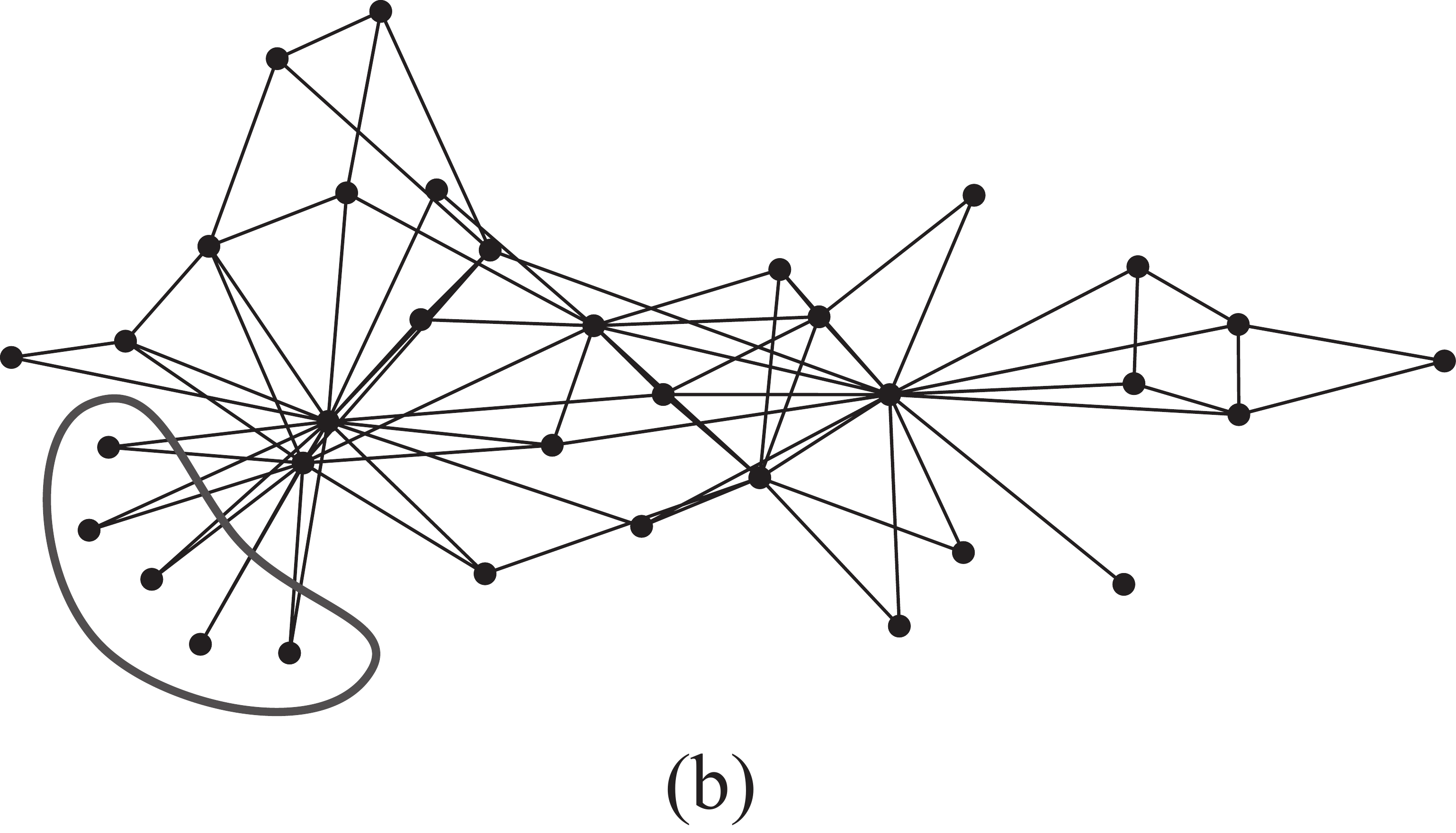}
\caption{(a) A hierarchical network; (b)  Zachary's karate-club network~\cite{zk}.
Encircled nodes have the structures in Fig.~\ref{trois}(b).}
\label{networksamples}
\end{figure}

\begin{figure}
\centering
\includegraphics[width=0.45\textwidth]{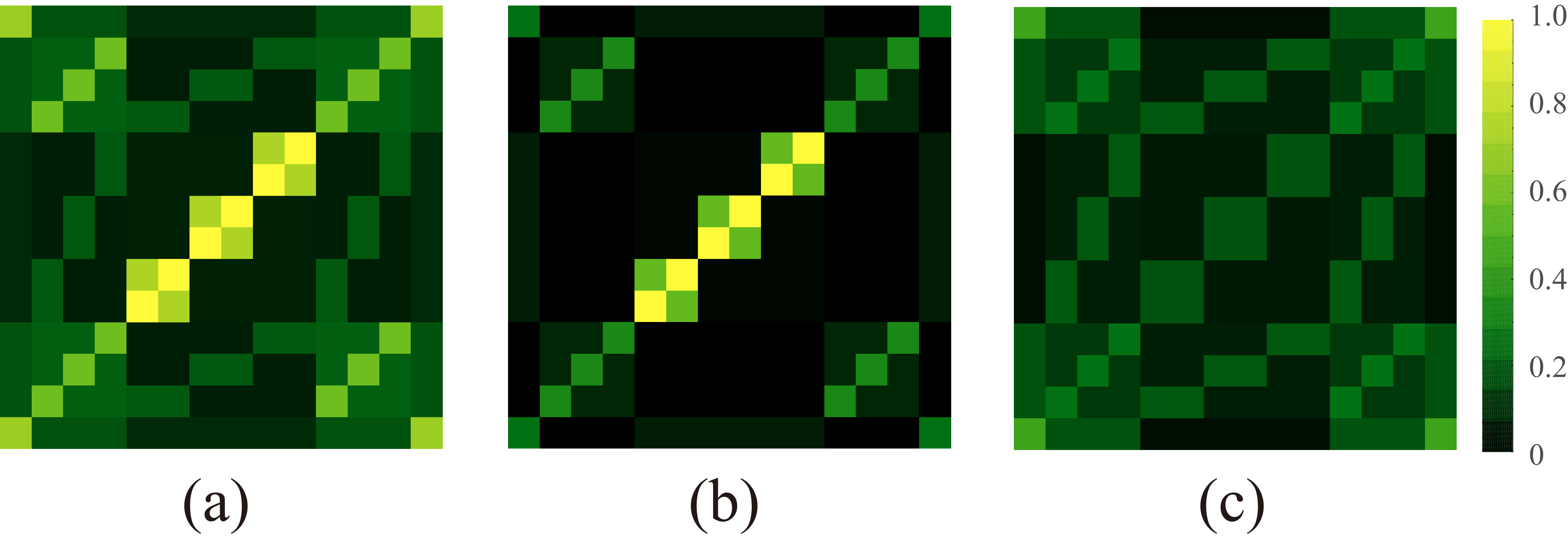}
\\
\vspace{\baselineskip}
\centering
\includegraphics[width=0.45\textwidth]{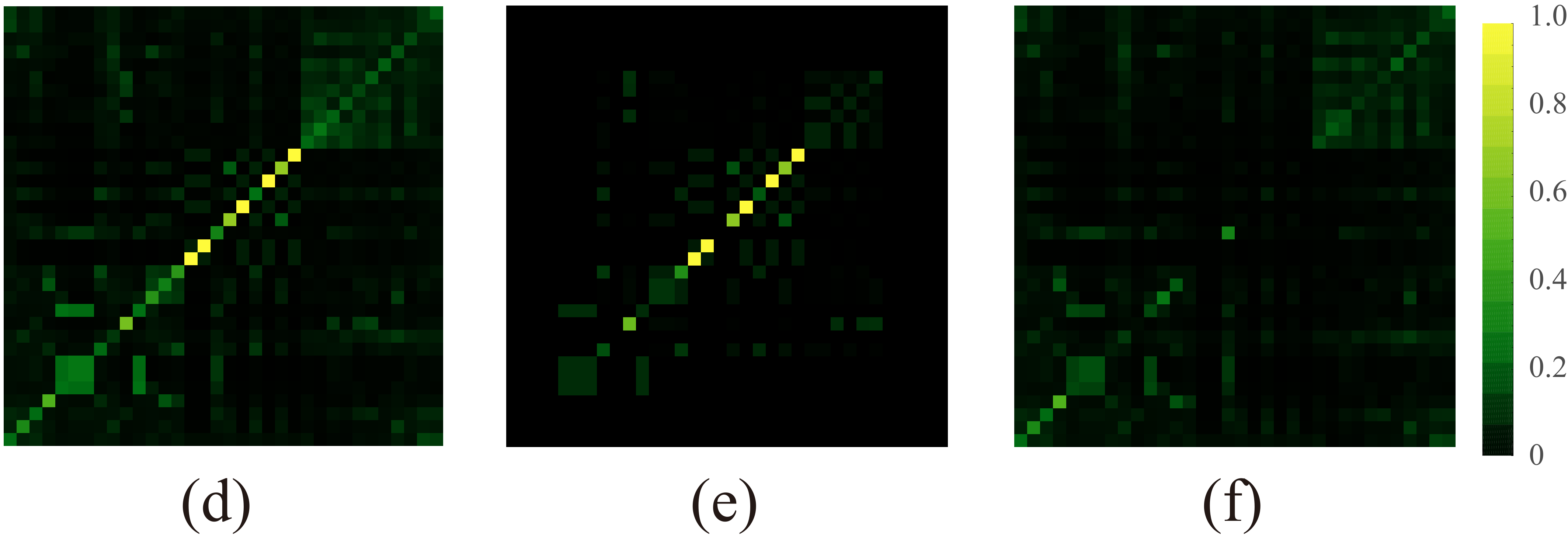}
\caption{Plots of $\chi_{jk}$ in Eq.~\eqref{eq70} (a--c) for the hierarchical network in Fig.~\ref{networksamples}(a)  and (d--f) for Zachary's karate club in Fig.~\ref{networksamples}(b). In each row, (a) and (d) represent the total of $\chi_{jk}$, (b) and (e) the contribution of the null eigenspace, $\chi_{jk}^{(0)}$, and (c) and (f) the remnants, $\chi_{jk}-\chi_{jk}^{(0)}$.
The horizontal and vertical axes indicate $k$ and $j$, respectively.
The legend should be augmented by a factor $\times 3$ in (a--c) and $\times 3/2$ in (d--f). 
}
\label{ctsplit}
\end{figure}

We now show that the quantum particle is localized on the nodes encircled in Fig.~\ref{networksamples}, which have the structure of Fig.~\ref{trois}(b), and that it is mainly due to the dominant contribution of the null eigenvalues. 
Figure~\ref{ctsplit} presents the density plots of $\chi_{jk}$ for Fig.~\ref{networksamples}.

A hierarchical network in Fig.\ref{networksamples}~(a), which we introduce here for explanatory purposes, has six null eigenvalues out of $N=14$ (that is, $N_0=6$ so that $N_0(N_0-1)=30>N$), which are generated by the three local structures encircled. 
Zachary's karate-club network in Fig.~\ref{networksamples}(b) has ten null eigenvalues out of $N=34$ (that is, $N_0=10$ and $N_0(N_0-1)=90>N$) due to five local structures as in Fig.~\ref{trois}(b) formed by the nodes encircled.

The panels (a) and (d) in Fig.~\ref{ctsplit} show $\chi_{jk}$ in Eq.~\eqref{eq70}, while the panels (b) and (e) show the contributions from the null eigenspace, 
\begin{align}
\chi_{jk}^{(0)}=2 \sum_{\shortstack{$\scriptstyle \mu<\nu$ \\ $\scriptstyle \lambda_\mu=\lambda_\nu=0$}} \langle j | \phi_\mu \rangle \langle \phi_\mu | k \rangle \langle k | \phi_\nu \rangle \langle \phi_\nu | j \rangle,
\end{align}
which comes from the second term of Eq.~\eqref{eq70} for the null eigenvalues $\lambda_\mu=\lambda_\nu=0$.
where
The panels (c) and (f) show the remnants, $\chi_{jk}-\chi_{jk}^{(0)}$.
We can see that the main part of $\chi_{jk}$ comes from the null-eigenvalue contribution $\chi_{jk}^{(0)}$. This means that the quantum particle stays on the encircled nodes in Fig.~\ref{networksamples}, namely the local structures in Fig.~\ref{trois}(b), and this localization is due to the null eigenspace.
This is another instance of the null-eigenvalue localization, which concept we introduced in the previous section. 

It is interesting to note that in the hierarchical network of Fig.~\ref{networksamples}(a), the CTQW is quite localized on other nodes too, as shown in Fig.~\ref{ctsplit}; these are due to the second condition of the three listed above, namely the partial duplication.  For example, the rows for the motifs \{2, 5, 6, 11\} are equal to the motif \{3, 7, 8, 12\}. We can therefore consider that the CTQW is localized, not on one node but on a full motif, which is exemplified by the high probabilities on $\chi_{56}$ and $\chi_{78}$ in Fig.~\ref{ctsplit}(a).

This leads us to the following. The null-eigenvalue localization happens in the case of both complete and partial duplications of network nodes. In the former case, the quantum particle is localized on two nodes, while in the latter, they are localized on a set of nodes. The more nodes are involved in the initial state of the partial duplication, the more the initial states is spread, and so the less we can consider it localized. 

\section{Conclusions}
\label{sec4}
In the present paper we analyzed the systematically abundant nullity graphs of complex networks in the context of quantum mechanics. 
These null eigenvalues are due to duplication mechanisms and local structures. 
The IPR and the CTQW revealed that the corresponding eigenstates are strongly localized on these structures.
We note that the latter should be more tractable than the former if one plans to detect the null-eigenvalue localization in some experimental situations, because precise tuning to the eigenvalue zero is not necessary.

In order to grow such structures, we need to place links in a correlated way, as opposed to a random one~\cite{growth}. This observation implies that connections are ultimately due to the growing mechanism of the preferential attachment developed by Barab\'{a}si and Albert~\cite{scalefree1,scalefree1-5}, which means that the higher the degree of a node is, the higher the probability of receiving new links will be.

We stress here that physics behind the null-eigenvalue localization is essentially different from the Anderson localization~\cite{localander,localization1} in two ways:
first, the eigenvalues of the latter lie on the edges of the density of states, whereas here they lie at its center;
second, the eigenstates of the Anderson localization typically decays exponentially, whereas here they are strictly caged in the local structures.
A similar remark was highlighted in Ref.~\cite{alloy} for regular lattices.
The Anderson localization is due to breaking of symmetries by means of randomness, 
while the null-eigenvalue localization is due to symmetries in the network structure, and is more similar to the bound state in continuum~\cite{BIC1,BIC2,Ordonez06,Tanaka07}.

Knowing this can give new insights on the design of quantum systems. For instance, if we want the particle to spread all over the network, we could argue that the system should be as random as possible in order to avoid local symmetries. 
This is consistent with Ref.~\cite{enhancing}, in which the authors proved numerically that breaking symmetries enhance spreading of a quantum walker. On the other hand, if we want to localize the particle, we should organize the nodes according to our study. 

One possibility of finding the present situation in reality is a chemical reaction chain in metabolism, such as photosynthesis in plants. Oxidation and reduction occur when electrons hop from one chemical to another. It has been documented that chemical reactions connect molecules to form a complex network, e.g.\ in the protein interaction network~\cite{protein-network1,protein-network2}. In experiments, quantum walks have been simulated in optics~\cite{quantum-walk-experiment1,quantum-walk-experiment2}. It may be possible to realize the present zero-eigenvalue localization experimentally in artificial optical devices. 
We could even imagine a situation in which we can control the correlation of electrons inserted in the network and understand how the degeneracies would impact on their propagation, similarly to Ref.~\cite{darkstate}, and in line with our understanding of partial duplications. 

Localization phenomena were previously reported in the case of the discrete-time quantum walk~\cite{DiscreteLoca1, DiscreteLoca2,Mukai20}. A comparative analysis of both cases would be useful to characterize this phenomenon better. Generalization to open quantum systems can also be interesting, for instance by analyzing the impact of the degeneracy when we add a lead to the system. Another valuable study would be to consider the localization patterns of various graphs and comparing the numbers of complete and partial duplications. A particular graph of interest for such work would be fractal graphs. Certain networks in nature that require strong localization may have a growth mechanism which generates significant duplications, and vice versa. A recent study~\cite{neural} even suggested a possible connection between peaks in the density of states of neural networks and the cognitive functionality; the connection might be indeed due to the localization of signals in neural networks.

\section{Acknowledgments}

N.H.'s work is supported by the following fundings: (i) the Aihara Innovative Mathematical Modelling Project, the Japan Society for the Promotion of Science (JSPS) through the ``Funding Program for World-Leading Innovative R\&D on Science and Technology (FIRST Program)," initiated by the Council for Science and Technology Policy (CSTP); (ii) the Impulsing Paradigm Change through Disruptive Technologies (ImPACT) Program of the Council of Science, Technology and Innovation (Cabinet Office, Government of Japan); (iii) the JSPS grant-in-aid (KAKENHI) No.~JP19H00658.

\end{document}